\documentclass[aps,prl,twocolumn,superscriptaddress,groupedaddress]{revtex4}  
\usepackage{graphicx}  
\usepackage{dcolumn}   
\usepackage{bm}        
\usepackage{amssymb}   

\begin{document}



\title{Study of baryon acoustic oscillations with SDSS DR12 data
and measurements of $\Omega_k$ and $\Omega_\textrm{DE}(a)$. Part II.}
\author{B.~Hoeneisen} \affiliation{Universidad San Francisco de Quito, Quito, Ecuador}

\date{\today}

\begin{abstract}
\noindent
We define Baryon Acoustic Oscillation (BAO) observables
$\hat{d}_\alpha(z, z_c)$, $\hat{d}_z(z, z_c)$, 
and $\hat{d}_/(z, z_c)$ that do not
depend on any cosmological parameter.
From each of these observables we recover the
BAO correlation length $d_\textrm{BAO}$ with its
respective dependence on cosmological parameters.
These BAO observables
are measured as a function of redshift $z$ with the
Sloan Digital Sky Survey (SDSS) data release DR12.
From the BAO measurements alone, or together
with the correlation angle $\theta_\textrm{MC}$
of the Cosmic Microwave Background (CMB), we constrain
the curvature parameter
$\Omega_k$ and the dark energy density
$\Omega_\textrm{DE}(a)$ as a function of the 
expansion parameter $a$ in several scenarios.
These observables are further constrained with external	
measurements of	$h$ and	$\Omega_\textrm{b} h^2$.
We find some tension between the                              
data and a cosmology with flat space and constant dark energy
density $\Omega_\textrm{DE}(a)$.
\end{abstract}

\maketitle

\section{Introduction}

We present studies of baryon acoustic oscillations (BAO)
with Sloan Digital Sky Survey (SDSS) data release DR12 \cite{DR12}.
Part I of this ongoing study is Ref. \cite{bh1}.
We refer the reader to Part I for the notation and methods.
This is an outline of the study:

(i) We define and measure BAO observables
$\hat{d}_\alpha(z, z_c)$, $\hat{d}_z(z, z_c)$, and $\hat{d}_/(z, z_c)$
that do	not depend on any cosmological parameter, see Part I \cite{bh1}.
From each of these observables we obtain
the BAO	correlation distance $d_\textrm{BAO}$ 
with its respective dependence on the cosmological parameters.
We present many redundant measurements with different
galaxy selections, i.e. different background fluctuations,
to gain confidence in the results.

(ii) We use the measured BAO distances, and the
correlation angle $\theta_\textrm{MC}$ of the
Cosmic Microwave Background (CMB), as an   
uncalibrated standard ruler to constrain the
correlated parameters $\Omega_k$ and $\Omega_\textrm{DE}(a)$.
The cosmological parameters $h$, $\Omega_\textrm{b} h^2$
and $N_\textrm{eff}$ drop out of this analysis.

(iii) Finally we use the measured BAO distances and $\theta_\textrm{MC}$
as a calibrated standard ruler to
further	constrain $\Omega_k$ with external measurements
of $h$ and $\Omega_\textrm{b} h^2$.

The results of this analysis are compared with
the final consensus results corresponding to the
DR12 data \cite{consensus}.

\begin{table*}
\caption{\label{cc}
Measured BAO distances $\hat{d}_\alpha(z, z_c)$,
$\hat{d}_z(z, z_c)$, and $\hat{d}_/(z, z_c)$ in units of $c/H_0$
with $z_c = 3.79$ (see Part I \cite{bh1}) from SDSS DR12 galaxies with
right ascension $110^0$ to $270^0$, and declination $-5^0$ to $70^0$
in the northern galactic cap, i.e. 
dec $> 27.0^0 - 17.0^0 \left[ (\textrm{ra} - 185.0^0)/(260.0^0 - 185.0^0) \right]^2$.
Uncertainties are statistical from the fits to the BAO signal.
Each BAO distance has an independent systematic uncertainty $\pm 0.00085$.
No corrections have been applied.
}
\begin{ruledtabular}
\begin{tabular}{c|ccccc|ccc}
$z$ & $z_\textrm{min}$ & $z_\textrm{max}$ & galaxies & centers & type &
$100 \hat{d}_\alpha(z, z_c)$ & $100 \hat{d}_z(z, z_c)$ & $100 \hat{d}_/(z, z_c)$ \\
\hline
$0.10$ & 0.0 & 0.2 & 187485 & 187485 & G-G &  &  &  \\
$0.10$ & 0.0 & 0.2 & 184044 & 27485 & G-C & $3.434 \pm 0.008$ &  & $3.340 \pm 0.013$ \\
$0.25$ & 0.2 & 0.3 & 38672 & 38672 & G-G & & & $3.369 \pm 0.012$ \\
$0.25$ & 0.2 & 0.3 & 38657 & 22079 & G-C & $3.284 \pm 0.013$ & $3.223 \pm 0.013$ & $3.334 \pm 0.009$ \\
$0.35$ & 0.3 & 0.4 & 53462 & 53462 & G-G & $3.246 \pm 0.016$ & $3.210 \pm 0.013$ & $3.295 \pm 0.009$ \\
$0.35$ & 0.3 & 0.4 & 53461 & 24360 & G-C & $3.324 \pm 0.009$ & $3.376 \pm 0.015$ & $3.288 \pm 0.010$ \\
$0.46$ & 0.4 & 0.5 & 86528 & 86528 & G-G & $3.538 \pm 0.009$ &  & $3.483 \pm 0.023$ \\
$0.46$ & 0.4 & 0.5 & 86528 & 21379 & G-C & $3.487 \pm 0.004$ & $3.319 \pm 0.011$ & $3.458 \pm 0.015$ \\
$0.54$ & 0.5 & 0.6 & 116301 & 116301 & G-G & $3.380 \pm 0.010$ & $3.486 \pm 0.011$ & $3.583 \pm 0.013$ \\
$0.54$ & 0.5 & 0.6 & 116301 & 19901 & G-C & $3.460 \pm 0.014$ & $3.471 \pm 0.014$ & $3.509 \pm 0.014$ \\
$0.67$ & 0.6 & 0.9 & 67772 & 67772 & G-G & $3.360 \pm 0.030$ & & $3.451 \pm 0.013$ \\
$0.67$ & 0.6 & 0.9 & 67772 & 3800 & G-C & $3.273 \pm 0.012$ & & $3.492 \pm 0.015$ \\
\hline
0.20 & 0.10 & 0.35 & 160471 & 160471 & G-G & & & \\
0.20 & 0.10 & 0.35 & 160471 & 59392 & G-C & & & $3.295 \pm 0.011$ \\
0.47 & 0.35 & 0.55 & 178246 & 178246 & G-G & $3.419 \pm 0.017$ & $3.285 \pm 0.015$ & $3.486 \pm 0.010$ \\
0.46 & 0.35 & 0.55 & 178246 & 45592 & G-C & $3.489 \pm 0.008$ & $3.350 \pm 0.010$ & $3.462 \pm 0.008$ \\
0.63 & 0.55 & 0.90 & 119449 & 119449 & G-G & & & $3.512 \pm 0.022$ \\
0.63 & 0.55 & 0.90 & 119449 & 12003 & G-C & & & \\
\end{tabular}
\end{ruledtabular}
\end{table*}

\begin{table*}
\caption{\label{ss}
Measured BAO distances $\hat{d}_\alpha(z, z_c)$,
$\hat{d}_z(z, z_c)$, and $\hat{d}_/(z, z_c)$ in units of $c/H_0$
with $z_c = 3.79$ (see Part I \cite{bh1}) from SDSS DR12 galaxies with
right ascension $110^0$ to $270^0$, and declination $-5^0$ to $70^0$
in the southern galactic cap, i.e.
dec $< 27.0^0 - 17.0^0 \left[ (\textrm{ra} - 185.0^0)/(260.0^0 - 185.0^0) \right]^2$.
Uncertainties are statistical from the fits to the BAO signal.
Each BAO distance has an independent systematic uncertainty $\pm 0.00085$.
No corrections have been applied.
}
\begin{ruledtabular}
\begin{tabular}{c|ccccc|ccc}
$z$ & $z_\textrm{min}$ & $z_\textrm{max}$ & galaxies & centers & type &
$100 \hat{d}_\alpha(z, z_c)$ & $100 \hat{d}_z(z, z_c)$ & $100 \hat{d}_/(z, z_c)$ \\
\hline
$0.10$ & 0.0 & 0.2 & 103008 & 103008 & G-G &  &  &  \\
$0.10$ & 0.0 & 0.2 & 117786 & 14802 & G-C & $3.262 \pm 0.009$ & $3.350 \pm 0.013$ & $3.351 \pm 0.003$ \\
$0.25$ & 0.2 & 0.3 & 21112 & 21112 & G-G & $3.420 \pm 0.019$ & $3.307 \pm 0.008$ & $3.529 \pm 0.010$ \\
$0.25$ & 0.2 & 0.3 & 21108 & 11320 & G-C & $3.473 \pm 0.012$ & $3.292 \pm 0.010$ & $3.504 \pm 0.009$ \\
$0.35$ & 0.3 & 0.4 & 29739 & 29739 & G-G & $3.408 \pm 0.032$ & $3.347 \pm 0.016$ & $3.464 \pm 0.031$ \\
$0.35$ & 0.3 & 0.4 & 29739 & 12907 & G-C & $3.388 \pm 0.010$ & $3.327 \pm 0.010$ & $3.463 \pm 0.011$ \\
$0.46$ & 0.4 & 0.5 & 46447 & 46447 & G-G & $3.501 \pm 0.015$ &  & $3.280 \pm 0.027$ \\
$0.46$ & 0.4 & 0.5 & 46447 & 9410 & G-C &  & & \\
$0.54$ & 0.5 & 0.6 & 65217 & 65217 & G-G & $3.302 \pm 0.022$ & $3.375 \pm 0.027$ & $3.416 \pm 0.016$ \\
$0.54$ & 0.5 & 0.6 & 65217 & 9943 & G-C &  &  &  \\
$0.67$ & 0.6 & 0.9 & 35100 & 35100 & G-G & $3.438 \pm 0.014$ &  & $3.399 \pm 0.018$ \\
$0.67$ & 0.6 & 0.9 & 35100 & 1494 & G-C &  &  &  \\
\hline
0.19 & 0.10 & 0.35 & 92687 & 92687 & G-G & & & \\
0.19 & 0.10 & 0.35 & 92687 & 30798 & G-C & $3.338 \pm 0.010$ & & \\
0.47 & 0.35 & 0.55 & 98467 & 98467 & G-G & & & $3.481 \pm 0.015$ \\
0.47 & 0.35 & 0.55 & 98467 & 22063 & G-C & $3.604 \pm 0.008$ & $3.357 \pm 0.023$ & $3.497 \pm 0.012$ \\
0.62 & 0.55 & 0.90 & 63441 & 63441 & G-G & $3.326 \pm 0.022$ & & $3.384 \pm 0.016$ \\
0.62 & 0.55 & 0.90 & 63441 & 5389 & G-C & & & $3.406 \pm 0.009$ \\
\end{tabular}
\end{ruledtabular}
\end{table*}

\section{BAO distances in the northern and southern galactic caps}

To gain confidence in the identification of the
BAO signal from among the background fluctuations,
and to obtain the uncertainties by a different
method, we repeat the measurements separately
for galaxies in the northern and southern
galactic caps. This time we analyze 
SDSS DR12 galaxies (passing the same quality 
selection flags and zErr $< 0.001$) with
right ascension $110^0$ to $270^0$ and declination $-5^0$ to $70^0$
with $17.0 < r_{35} < 27.0$. The galactic plane separates this
sample into two independent sub-samples defined by
dec $\gtrless 27.0^0 - 17.0^0 \left[ (ra - 185.0^0)/(260.0^0 - 185.0^0) \right]^2$.
These sub-samples are referred to as northern ($>$) and southern ($<$)
galactic caps. The results of measurements for G-G and G-C runs
are presented in Tables \ref{cc} and \ref{ss}. Blank entries indicate
that we were unable to reliably identify a BAO signal.

We combine the
measurements of the two independent sub-samples as follows: 
(i) if fits are successful for both G-G and G-C runs we take
the arithmetic average of the two measurements;
(ii) we then take the arithmetic average of each 
corresponding measurement
for the northern and southern galactic caps, and
assign to each of these averages an
independent total uncertainty $\sigma$ equal to 
half the total root-mean-square (r.m.s.) difference of all measurements in 
the northern and southern galactic caps; and finally,
(iv) assign an independent systematic 
uncertainty $\sqrt{2} \sigma$ to each
entry in Tables \ref{cc} and \ref{ss}. This procedure 
obtains almost the same uncertainties
as the method used for Table III of Part I \cite{bh1} in spite of the
fact that there are more galaxies in the northern galactic
cap than in the southern galactic cap, 
i.e. we find larger background fluctuations in the north.
The averages are summarized in Table \ref{cs}.

The two sets of measurements presented in Table III of 
Part I \cite{bh1} and Table \ref{cs}
have different background fluctuations due to galaxy clustering,
have different fits, obtain uncertainties by different
methods, and obtain essentially the same results and uncertainties.
The root-mean-square of the differences of all entries in
Table III of Part I \cite{bh1} and Table \ref{cs} divided by $\sqrt{2}$ is 0.00046
which is less than the total independent 
uncertainties assigned to
each entry of each Table, so these measurements are
consistent.

\begin{table}
\caption{\label{cs}
Independent measured BAO distances $\hat{d}_\alpha(z, z_c)$,
$\hat{d}_z(z, z_c)$, and $\hat{d}_/(z, z_c)$ in units of $c/H_0$
with $z_c = 3.79$ (see Part I \cite{bh1}) obtained by averaging measurements 
in the northern and southern galactic caps.
Each BAO distance has an independent total uncertainty $\pm 0.00060$
dominated by systematics.
No corrections have been applied.
}
\begin{ruledtabular}
\begin{tabular}{c|cc|ccc}
$z$ & $z_\textrm{min}$ & $z_\textrm{max}$ &
$100 \hat{d}_\alpha(z, z_c)$ & $100 \hat{d}_z(z, z_c)$ & $100 \hat{d}_/(z, z_c)$ \\
\hline
$0.10$ & 0.0 & 0.2 & $3.348$ & $3.350$ & $3.346$ \\
$0.25$ & 0.2 & 0.3 & $3.365$ & $3.261$ & $3.434$ \\
$0.35$ & 0.3 & 0.4 & $3.342$ & $3.315$ & $3.378$ \\
$0.46$ & 0.4 & 0.5 & $3.507$ & $3.319$ & $3.375$ \\
$0.54$ & 0.5 & 0.6 & $3.361$ & $3.427$ & $3.481$ \\
$0.67$ & 0.6 & 0.9 & $3.377$ &         & $3.435$ \\
\end{tabular}
\end{ruledtabular}
\end{table}

\section{Corrections}

Let us consider	corrections to the BAO distances 
due to peculiar	velocities and peculiar displacements
of galaxies towards their centers.
A relative peculiar velocity $v_p$ towards the center
causes a reduction of the BAO distances
$\hat{d}_\alpha(z, z_c)$, $\hat{d}_z(z, z_c)$, and $\hat{d}_/(z, z_c)$
of order $0.5 v_p/c$. In addition, the Doppler shift
produces an apparent shortening of $\hat{d}_z(z, z_c)$
by $v_p/c$, and somewhat less for $\hat{d}_/(z, z_c)$.

We multiply the measured BAO distances
$\hat{d}_\alpha(z, z_c)$, $\hat{d}_z(z, z_c)$, and $\hat{d}_/(z, z_c)$
by correction factors $f_\alpha$, $f_z$ and $f_/$ respectively.
Simulations in Ref. \cite{Seo} obtain
$f_\alpha - 1 = 0.2283 \pm 0.0609 \%$ and $f_z - 1 = 0.2661 \pm 0.0820 \%$
at $z = 0.3$, 
$f_\alpha - 1 = 0.1286 \pm 0.0425 \%$ and $f_z - 1 = 0.1585 \pm 0.0611 \%$
at $z = 1$, and
$f_\alpha - 1 = 0.0435 \pm 0.0293 \%$ and $f_z - 1 = 0.0582 \pm 0.0402 \%$
at $z = 3$.
In the following sections we present fits with the
corrections
\begin{eqnarray}
f_\alpha - 1 & = & 0.320\% a^{1.35}, \nonumber \\ 
f_z - 1 & = & 0.381\% a^{1.35}, \nonumber \\
f_/ - 1 & = & 0.350\% a^{1.35}.
\label{correction}
\end{eqnarray}

The effect of these corrections can be seen by comparing
the first two fits in Table \ref{BAO_fit} below. Fits with 
corrections $\sim 15$ times larger,
or no corrections at all, are presented in Part I \cite{bh1}.
An order-of-magnitude estimate of this correction
can be obtained by calculating the r.m.s. 
$v_p$ corresponding to modes with 
$k \equiv 2 \pi / \lambda < 2 \pi / (4 d'_\textrm{BAO})$
with Eq. (11) of Ref. \cite{BH} and normalizing the result to 
$\sigma_8$, i.e. to the r.m.s. density fluctuation in a volume
$(8 \textrm{Mpc}/h)^3$.

\begin{table*}
\caption{\label{dr12}
Final consensus ``BAO$+$FS" measurements of the	DR12 data set \cite{consensus}
(uncertainties are statistical and systematic), 
and the	corresponding BAO parameters $\hat{d}_\alpha(z, z_c)$ and
$\hat{d}_z(z, z_c)$ with $z_c =	3.79$. These measurements include the
peculiar motion	corrections. For comparison, the quoted errors 
on $\hat{d}_\alpha(z, z_c)$ and $\hat{d}_z(z, z_c)$ exclude the uncertainty
of $H_0$. Also shown is $\Omega_\textrm{DE}(z)$ extracted from $H$ with
$\Omega_k = 0$ and $\Omega_\textrm{m} = 0.310 \pm 0.005$ 
\cite{consensus} with
uncorrelated and correlated uncertainties.
}
\begin{ruledtabular}
\begin{tabular}{c|cc|ccc}
$z$ & $D_M r_\textrm{d,fid} / r_\textrm{d}$ [Mpc] & $100 \hat{d}_\alpha(z, z_c)$ &
 $H r_\textrm{d} / r_\textrm{d,fid} [\textrm{km s}^{-1} \textrm{Mpc}^{-1}]$ &
 $100 \hat{d}_z(z, z_c)$ & $\Omega_\textrm{DE}(z)$ \\
\hline
0.38 & $1518 \pm 20 \pm 11$ & $3.346 \pm 0.050$ 
 & $81.5 \pm 1.7 \pm 0.9$ & $3.270 \pm 0.077$ 
 & $0.698 \pm 0.071 \textrm{ (uncorr)} \pm 0.058 \textrm{ (corr)}$ \\
0.51 & $1977 \pm 23 \pm 14$ & $3.332 \pm 0.045$ 
 & $90.5 \pm 1.7 \pm 1.0$ & $3.375 \pm 0.074$
 & $0.798 \pm 0.081 \textrm{ (uncorr)} \pm 0.070 \textrm{ (corr)}$ \\
0.61 & $2283 \pm 28 \pm 16$ & $3.362 \pm 0.047$ 
 & $97.3 \pm 1.8 \pm 1.1$ & $3.426 \pm 0.061$
 & $0.862 \pm 0.077 \textrm{ (uncorr)} \pm 0.081 \textrm{ (corr)}$ 
\end{tabular}
\end{ruledtabular}
\end{table*}

\begin{table*}
\caption{\label{BAO_fit}
Cosmological parameters obtained from the 17 BAO measurements in Table \ref{cs}
in several scenarios. Corrections for peculiar motions are given by
Eq. (\ref{correction}) except, for comparison, the fit ``1*" which has no correction.
Scenario 1 has $\Omega_\textrm{DE}(a)$ constant.
Scenario 3 has $w = w_0$.
Scenario 4 has $\Omega_\textrm{DE}(a) = \Omega_\textrm{DE} \left[1 + w_1 (1 - a)\right]$.
}
\begin{ruledtabular}
\begin{tabular}{c|cccccc} 
   & Scenario 1* & Scenario 1 & Scenario 1 & Scenario 3 & Scenario 4 & Scenario 4 \\
\hline
$\Omega_k$ & $0$ fixed  & $0$ fixed  & $-0.413 \pm 0.234$ & $0$ fixed & $0$ fixed 
  & $-0.377 \pm 0.263$ \\
$\Omega_\textrm{DE} + 0.6 \Omega_k$ & $0.680 \pm 0.023$ & $0.682 \pm 0.023$ & $0.647 \pm 0.031$ 
  & $0.614 \pm 0.051$ & $0.611 \pm 0.073$ & $0.627 \pm 0.075$ \\
$w_0$ & n.a. & n.a. & n.a. & $-1.346 \pm 0.349$ & n.a. & n.a. \\
$w_a$ or $w_1$ & n.a. & n.a. & n.a. & n.a. & $-1.075 \pm 1.159$ & $-0.244 \pm 0.894$ \\
$100 d_\textrm{BAO}$ & $3.37 \pm 0.03$ & $3.38 \pm 0.03$ & $3.40 \pm 0.03$ 
  & $3.44 \pm 0.07$ & $3.44 \pm 0.07$ & $3.42 \pm 0.07$ \\
$\chi^2/$d.f. & $15.6/15$ & $15.4/15$ & $12.5/14$ & $14.3/14$ & $14.3/14$ & $12.4/13$ \\
\end{tabular}
\end{ruledtabular}
\end{table*}

\begin{table*}
\caption{\label{BAO_thetaMC_fit_18}
Cosmological parameters obtained from the 18 BAO measurements 
in Table III of Part I \cite{bh1}
plus $\theta_\textrm{MC}$ in several scenarios.
Corrections for peculiar motions are given by
Eq. (\ref{correction}).
Scenario 1 has $\Omega_\textrm{DE}(a)$ constant.
Scenario 2 has $w(a) = w_0 + w_a (1 - a)$.
Scenario 3 has $w = w_0$.
Scenario 4 has $\Omega_\textrm{DE}(a) = \Omega_\textrm{DE} \left[1 + w_1 (1 - a)\right]$.
}
\begin{ruledtabular}
\begin{tabular}{c|cccccc} 
   & Scenario 1 & Scenario 1 & Scenario 2 & Scenario 3 & Scenario 4 & Scenario 4 \\
\hline
$\Omega_k$ & $0$ fixed  & $0.037 \pm 0.011$ & $0$ fixed  & 
  $0$ fixed & $0$ fixed & $0.044 \pm 0.041$ \\
$\Omega_\textrm{DE} + 2 \Omega_k$ & $0.735 \pm 0.004$ & $0.718 \pm 0.006$ & $0.780 \pm 0.082$ 
  & $0.726 \pm 0.005$ & $0.720 \pm 0.006$ & $0.718 \pm 0.006$ \\
$w_0$ & n.a. & n.a. & $-0.904 \pm 0.107$ & $-0.815 \pm 0.051$ & n.a. & n.a. \\
$w_a$ or $w_1$ & n.a. & n.a. & $0.824 \pm 0.257$ & n.a. & $0.736 \pm 0.235$ & $-0.173 \pm 0.940$ \\
$100 d_\textrm{BAO}$ & $3.47 \pm 0.02$ & $3.39 \pm 0.03$ & $3.40 \pm 0.05$ 
  & $3.36 \pm 0.04$ & $3.35 \pm 0.04$ & $3.40 \pm 0.06$ \\
$\chi^2/$d.f. & $36.7/17$ & $23.6/16$ & $23.0/15$ & $24.0/16$ & $24.8/16$ & $23.6/15$ \\
\end{tabular}
\end{ruledtabular}
\end{table*}

\begin{table*}
\caption{\label{BAO_thetaMC_fit_17}
Cosmological parameters obtained from the 17 BAO measurements in Table \ref{cs}
plus $\theta_\textrm{MC}$ in several scenarios.
Corrections for peculiar motions are given by
Eq. (\ref{correction}).
Scenario 1 has $\Omega_\textrm{DE}(a)$ constant.
Scenario 2 has $w(a) = w_0 + w_a (1 - a)$.
Scenario 3 has $w = w_0$.
Scenario 4 has $\Omega_\textrm{DE}(a) = \Omega_\textrm{DE} \left[1 + w_1 (1 - a)\right]$.
}
\begin{ruledtabular}
\begin{tabular}{c|cccccc}
   & Scenario 1 & Scenario 1 & Scenario 2 & Scenario 3 & Scenario 4 & Scenario 4 \\
\hline
$\Omega_k$ & $0$ fixed  & $0.022 \pm 0.012$ & $0$ fixed & 
  $0$ fixed & $0$ fixed & $0.060 \pm 0.052$ \\
$\Omega_\textrm{DE} + 2 \Omega_k$ & $0.727 \pm 0.004$ & $0.716 \pm 0.007$ & $0.767 \pm 0.071$ 
  & $0.721 \pm 0.005$ & $0.719 \pm 0.006$ & $0.717 \pm 0.007$ \\
$w_0$ & n.a. & n.a. & $-0.997 \pm 0.103$ & $-0.895 \pm 0.056$ & n.a. & n.a. \\
$w_a$ or $w_1$ & n.a. & n.a. & $0.920 \pm 0.318$ & n.a. & $0.381 \pm 0.229$ & $-0.863 \pm 1.254$ \\
$100 d_\textrm{BAO}$ & $3.43 \pm 0.02$ & $3.38 \pm 0.03$ & $3.41 \pm 0.05$ 
  & $3.37 \pm 0.04$ & $3.37 \pm 0.04$ & $3.43 \pm 0.07$ \\
$\chi^2/$d.f. & $19.5/16$ & $15.7/15$ & $15.0/14$ & $16.1/15$ & $16.4/15$ & $15.1/14$ \\
\end{tabular}
\end{ruledtabular}
\end{table*}

\section{Comparison with the final consensus DR12 analysis}

We compare the measured BAO observables in Table III of Part I \cite{bh1} and Table \ref{cs}
with the final consensus ``BAO$+$FS" analysis of the DR12
data set \cite{consensus} which is summarized in Table \ref{dr12}.
The notation of Ref. \cite{consensus} is related to our notation
as follows:
\begin{eqnarray}
D_M \frac{r_\textrm{d,fid}}{r_\textrm{d}} & = &
 \frac{c}{H_0} \chi(z) \frac{r_\textrm{d,fid}}{d_\textrm{BAO}} \nonumber \\
 & = & \frac{c}{H_0} r_\textrm{d,fid} \frac{z \exp{(-z/z_c)}}{\hat{d}_\alpha(z, z_c)}, \\
H \frac{r_\textrm{d}}{r_\textrm{d,fid}} & = &
 H_0 E(z) \frac{d_\textrm{BAO}}{r_\textrm{d,fid}} \nonumber \\
 & = & \frac{H_0}{r_\textrm{d,fid}} \frac{\hat{d}_z(z, z_c)}{(1 - z/z_c) \exp{(-z/z_c)}},
\end{eqnarray}
where $r_\textrm{d,fid}	= 147.78$ Mpc and $H_0 = 67.8 \pm 1.2$ km s$^{-1}$ Mpc$^{-1}$.
We find agreement within the quoted uncertainties between our
measurements in Tables III of Part I \cite{bh1} or \ref{cs} and the
final consensus measurements in Table \ref{dr12}.

Table \ref{dr12}
also shows $\Omega_\textrm{DE}(z)$ extracted from $H$ with
$\Omega_k = 0$ and $\Omega_\textrm{m} = 0.310 \pm 0.005$ \cite{consensus}.
These values of $\Omega_\textrm{DE}(z)$ are in agreement with
our results in Fig. \ref{O_DE_z_all} below.
The observed increase of $\Omega_\textrm{DE}(z)$ with $z$ was 
studied in Part I \cite{bh1}.

\section{Constraints on $\Omega_k$ and $\Omega_\textrm{DE}(a)$
from uncalibrated BAO}

Let us try to understand qualitatively how 
the BAO distance measurements presented in 
Table \ref{cs} constrain the cosmological parameters.
In the limit $z \rightarrow 0$ we obtain
$d_\textrm{BAO} = \hat{d}_\alpha(0, z_c) = \hat{d}_z(0, z_c) = \hat{d}_/(0, z_c)$,
so the row with $z = 0.1$ in Table \ref{cs}
approximately determines $d_\textrm{BAO}$.
This $d_\textrm{BAO}$ and
the measurement of, for example, $\hat{d}_z(0.3, z_c)$
then constrains the derivative of 
$\Omega_\textrm{m}/a^3 + \Omega_\textrm{DE} + \Omega_\textrm{k}/a^2$
with respect to $a$ at $z \approx 0.3$, i.e.
constrains approximately 
$\Omega_\textrm{DE} + 0.5 \Omega_k$ or equivalently 
$\Omega_\textrm{DE} - \Omega_\textrm{m}$.
We need an additional constraint for Scenario 1.
At small $a$, $E(a)$ is dominated by $\Omega_\textrm{m}$,
so $\theta_\textrm{MC}$ plus $d_\textrm{BAO}$ approximately constrain
$\Omega_\textrm{m}$, or equivalently $\Omega_\textrm{DE} + \Omega_k$,
see Eq. (19) of Part I \cite{bh1}.
The additional BAO distance measurements in Table \ref{cs}
then also constrain $w_0$ and $w_a$ or $w_1$.

We now constrain cosmological parameters with each of these
three sets of independent BAO measurements:
18 BAO distances in Table III of Part I \cite{bh1},
12 BAO distances in Table IV of Part I (rows with $0.2 < z < 0.4$ G-C,
$0.4 < z < 0.5$ G-C, $0.5 < z < 0.6$ G-LC, and $0.6 < z < 0.9$ LG-LG),
or 17 BAO distances in Table \ref{cs}.

In Table \ref{BAO_fit} we present the
cosmological parameters obtained by minimizing the
$\chi^2$ with 17 terms corresponding to the
17 BAO distance measurements in Table \ref{cs} for
several scenarios. We find that the data is in
agreement with the simplest cosmology with
$\Omega_k = 0$ and $\Omega_\textrm{DE}(a)$ constant
with $\chi^2$ per degree of freedom (d.f.) $15.4/15$, so no additional
parameter is needed to obtain a good fit to this data.

For the sets of 18, 12 or 17 BAO measurements we obtain,
respectively,
$\Omega_\textrm{DE} + 0.6 \Omega_k = 0.620 \pm 0.030$,
$0.652 \pm 0.041$, and $0.647 \pm 0.031$ for constant
$\Omega_\textrm{DE}(a)$, or
$0.638 \pm 0.058$, $0.585 \pm 0.063$, and $0.627 \pm 0.075$ 
if $\Omega_\textrm{DE}(a)$ is allowed to
depend on $a$ as in Scenario 4.
We present the variable $\Omega_\textrm{DE} + 0.6 \Omega_k$
instead of $\Omega_\textrm{DE}$ because it has a smaller 
uncertainty.
The constraints on $\Omega_k$ are weak.

In Table \ref{BAO_thetaMC_fit_18} we present the
cosmological parameters obtained by minimizing the
$\chi^2$ with 19 terms corresponding to the
18 BAO distance measurements listed
in Table III of Part I \cite{bh1}
plus the correlation angle
$\theta_\textrm{MC} = 0.010413 \pm 0.000006$
of the CMB \cite{PDG}.
We present the variable $\Omega_\textrm{DE} + 2 \Omega_\textrm{k}$
instead of $\Omega_\textrm{DE}$ because it has a smaller uncertainty.
The corresponding fits for the
17 BAO measurements of Table \ref{cs}
plus $\theta_\textrm{MC}$ are presented in Table 
\ref{BAO_thetaMC_fit_17}.

From the fits for $\theta_\textrm{MC}$ plus the set of 18, or 12 or 17 BAO
measurements we obtain, respectively,
$\Omega_\textrm{DE} + 2 \Omega_\textrm{k} = 0.718 \pm 0.006$,
$0.749 \pm 0.024$, or $0.717 \pm 0.007$
when $\Omega_\textrm{DE}(a)$ is allowed to vary as in Scenario 4.
The constraints on $\Omega_k$ are, respectively,
$0.037 \pm 0.011$, $0.043 \pm 0.015$, or $0.022 \pm 0.012$
for constant $\Omega_\textrm{DE}(a)$, or
$0.044 \pm 0.041$, $0.116 \pm 0.057$, or $0.060 \pm 0.052$ 
when $\Omega_\textrm{DE}(a)$ is allowed to vary as in Scenario 4.
The constraints on $w_1$ are respectively
$0.74 \pm 0.24$, $1.00 \pm 0.40$, or $0.38 \pm 0.23$ 
for $\Omega_k = 0$.

Note that the BAO plus $\theta_\textrm{MC}$
data is consistent with $\Omega_k = 0$ or with constant
$\Omega_\textrm{DE}(a)$, i.e. $w_1 = 0$, but there is some tension
when both constraints are applied. 
A summary of tensions
is presented in Table \ref{tension}.
The tension is not statistically significant for the 17 BAO plus 
$\theta_\textrm{MC}$ data (in part because the 17 BAO set has
no measurement of $\hat{d}_z(0.67, z_c)$, see 
Figures \ref{O_DE_z} and \ref{O_DE_z_all} below).

\begin{table}
\caption{\label{tension}
There is some tension between the data and fits with $\Omega_k = 0$
and $\Omega_\textrm{DE}(a)$ constant,
i.e. $w_1 = 0$ in Scenario 4.
Shown is the 
reduction of the $\chi^2$ of the fits when either
$\Omega_k$ or $w_1$ is released.
The BAO measurements correspond to
Tables III or IV of Part I \cite{bh1}, or Table \ref{cs}.
Entries with * have no significant tension with 
$\Omega_k = 0$ and constant $\Omega_\textrm{DE}(a)$.
} 
\begin{ruledtabular}
\begin{tabular}{lrr}
Reduction of $\chi^2$ by releasing &   $\Omega_k$ & $w_1$ \\
\hline
Data: & & \\
18 BAO $+$ $\theta_\textrm{MC}$                &   13.1 &  11.9 \\
12 BAO $+$ $\theta_\textrm{MC}$                &   10.8 &   8.9 \\
17 BAO $+$ $\theta_\textrm{MC}$ *              &    3.8 &   3.1 \\
18 BAO $+$ $\theta_\textrm{MC}$ $+$ $A=1.000 \pm 0.022$ &  12.7 &  10.4 \\
12 BAO $+$ $\theta_\textrm{MC}$ $+$ $A=1.000 \pm 0.022$ &   9.8 &   6.9 \\
17 BAO $+$ $\theta_\textrm{MC}$ $+$ $A=1.000 \pm 0.022$ * &  4.0 &   2.3 \\
18 BAO $+$ $\theta_\textrm{MC}$ $+$ $A=0.968 \pm 0.012$ &   4.9 &  16.3 \\
12 BAO $+$ $\theta_\textrm{MC}$ $+$ $A=0.968 \pm 0.012$ &   2.2 &  16.7 \\
17 BAO $+$ $\theta_\textrm{MC}$ $+$ $A=0.968 \pm 0.012$ &   0.6 &  5.7 \\
\end{tabular}
\end{ruledtabular}
\end{table}

\section{Measurement of $\Omega_\textrm{DE}(a)$}

\begin{figure}
\begin{center}
\scalebox{0.45}
{\includegraphics{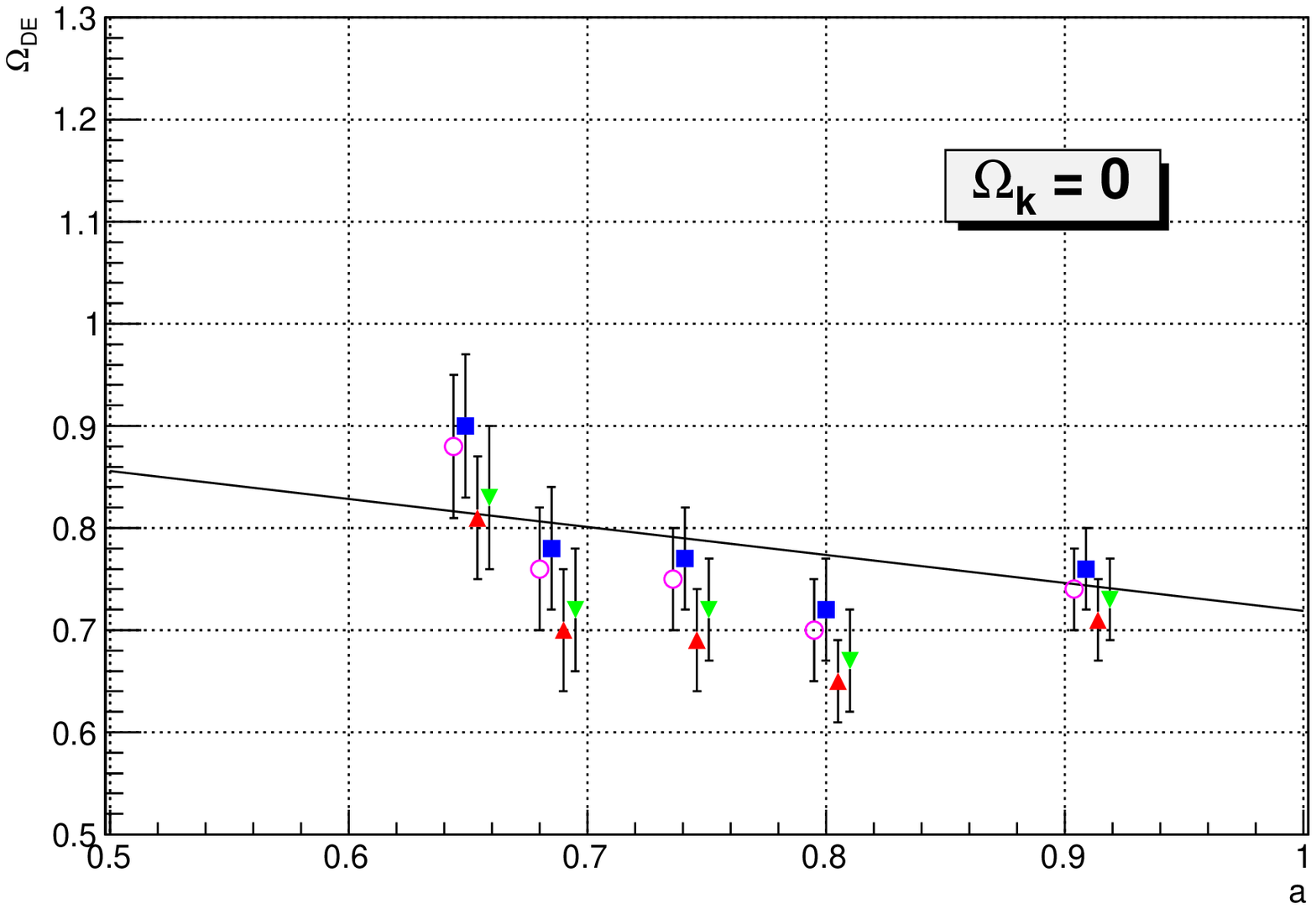}}
\caption{Measurements of
$\Omega_\textrm{DE}(a)$ obtained
from the 5 $\hat{d}_z(z, z_c)$ in Table \ref{cs} 
for $\Omega_k = 0$, and
the corresponding $d_\textrm{BAO}$ and $\Omega_\textrm{DE}$
from the fit for Scenario 4 in Table \ref{BAO_thetaMC_fit_17}.
The straight line is 
$\Omega_\textrm{DE}(a) = 0.7188 \left[ 1 + 0.3813 (1 - a) \right]$
from the central values of this fit.
The uncertainties correspond only to the total uncertainties
of $\hat{d}_z(z, z_c)$.
For clarity some offsets in $a$ have been applied.
We present results for
$(d_\textrm{BAO}, \Omega_\textrm{DE}) = (0.03367 - 0.00044, 0.7188)$ (squares),
$(0.03367 + 0.00044, 0.7188)$ (triangles),
$(0.03367, 0.7188 - 0.0063)$ (inverted triangles), and
$(0.03367, 0.7188 + 0.0063)$ (circles).
}
\label{O_DE_z}
\end{center}
\end{figure}

We obtain $\Omega_\textrm{DE}(a)$ from the 5 independent measurements of
$\hat{d}_z(z, z_c)$ in Table \ref{cs}
and Eqs. (17) and (2) of Part I \cite{bh1},
for the case $\Omega_k = 0$.
The values of $d_\textrm{BAO}$ and $\Omega_\textrm{m} = 1 - \Omega_\textrm{DE} - \Omega_k$
are obtained from the fit for Scenario 4 in Table \ref{BAO_thetaMC_fit_17}.
The results are presented in Fig. \ref{O_DE_z}. To guide the eye, we
also show the straight line corresponding to 
the central values of $\Omega_\textrm{DE}$ and $w_1$ of 
the fit for Scenario 4.

\begin{figure}
\begin{center}
\scalebox{0.45}
{\includegraphics{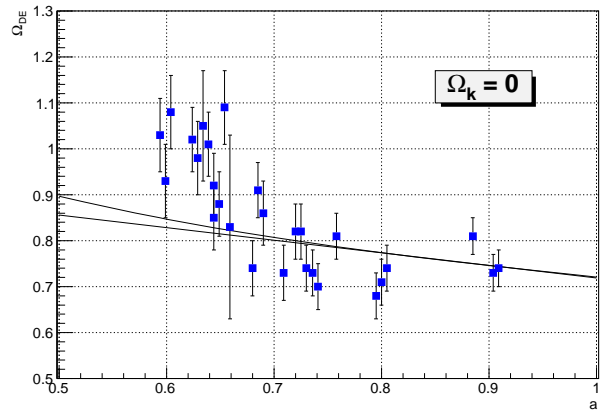}}
\caption{Same as Figure \ref{O_DE_z} with the addition of the 
6 measurements of $\hat{d}_z(z, z_c)$ in Table III of Part I \cite{bh1}, and
the 17 measurements of $\hat{d}_z(z, z_c)$ in Table IV of Part I.
These measurements are partially correlated.
We present results for
$(d_\textrm{BAO}, \Omega_\textrm{DE}) = (0.03367, 0.7188)$.
The curve corresponding to the fit for Scenario 3 in Table 
\ref{BAO_thetaMC_fit_17} has been added.
The excesses of $\Omega_\textrm{DE}(a)$ for $0.6 < a < 0.67$ 
are not understood. 
}
\label{O_DE_z_all}
\end{center}
\end{figure}

To check the robustness of $\Omega_\textrm{DE}(a)$ in Fig. \ref{O_DE_z}
we add the 6 measurements of $\hat{d}_z(z, z_c)$ in Table III of Part I \cite{bh1} and the
17 measurements of $\hat{d}_z(z, z_c)$ in Table IV of Part I and
obtain Fig. \ref{O_DE_z_all}. Note that these measurements of
$\hat{d}_z(z, z_c)$ are partially correlated.
Note that there is tension with a constant
$\Omega_\textrm{DE}(a)$ for $0.6 < a < 0.67$.
Note also that the final consensus measurements of DR12 \cite{consensus}
in the last column of Table \ref{dr12} are in agreement with
Fig. \ref{O_DE_z_all}, and also show the tension with
$\Omega_k = 0$ and constant $\Omega_\textrm{DE}(a)$.

\begin{figure}
\begin{center}
\scalebox{0.45}
{\includegraphics{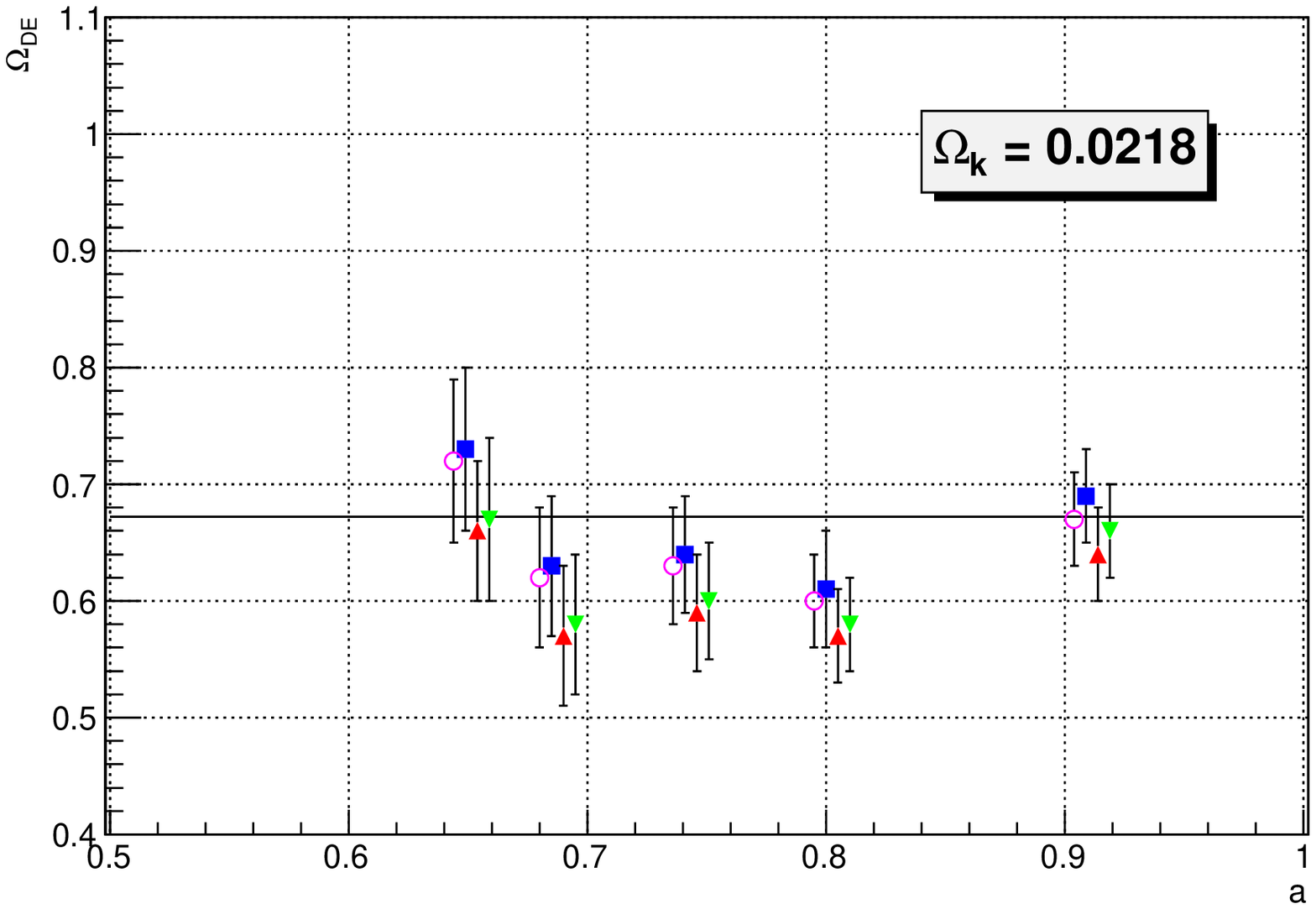}}
\caption{Measurements of
$\Omega_\textrm{DE}(a)$ obtained
from the 5 $\hat{d}_z(z, z_c)$ in Table \ref{cs} 
for $\Omega_k = 0.0218$, and
the corresponding $d_\textrm{BAO}$ and $\Omega_\textrm{DE}$
from the fit for Scenario 1 in Table \ref{BAO_thetaMC_fit_17}.
The straight line is 
$\Omega_\textrm{DE}(a) = 0.6723$ constant
from the central value of this fit.
The uncertainties correspond only to the total uncertainties
of $\hat{d}_z(z, z_c)$.
For clarity some offsets in $a$ have been applied.
We present results for
$(d_\textrm{BAO}, \Omega_\textrm{DE}) = (0.03383 - 0.00033, 0.6723)$ (squares),
$(0.03383 + 0.00033, 0.6723)$ (triangles),
$(0.03383, 0.6723 - 0.0068)$ (inverted triangles), and
$(0.03383, 0.6723 + 0.0068)$ (circles).
}
\label{O_DE_z_218}
\end{center}
\end{figure}

\begin{figure}
\begin{center}
\scalebox{0.45}
{\includegraphics{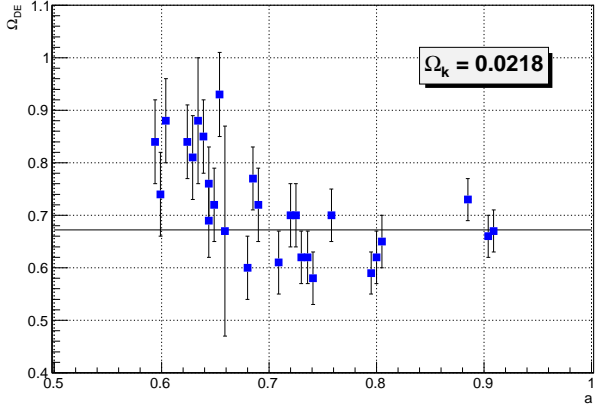}}
\caption{Same as Figure \ref{O_DE_z_218} with the addition of the 
6 measurements of $\hat{d}_z(z, z_c)$ in Table III of Part I \cite{bh1}, and
the 17 measurements of $\hat{d}_z(z, z_c)$ in Table IV of Part I.
These measurements are partially correlated.
We present results for
$(d_\textrm{BAO}, \Omega_\textrm{DE}) = (0.03383, 0.6723)$.
}
\label{O_DE_z_all_218}
\end{center}
\end{figure}

We repeat these two figures for the fit for Scenario 1 in
Table \ref{BAO_thetaMC_fit_17}, see Figs. \ref{O_DE_z_218} 
and \ref{O_DE_z_all_218}. For these figures $\Omega_k = 0.0218$.
The excesses of $\Omega_\textrm{DE}(a)$ for
$0.6 < a < 0.67$ are not understood.

\section{Constraints on $\Omega_k$ and $\Omega_\textrm{DE}(a)$
from calibrated BAO}

Up to this point we have used the BAO distance $d_\textrm{BAO}$
as an uncalibrated standard ruler.
The cosmological parameters $h$, $\Omega_\textrm{b} h^2$ and $N_\textrm{eff}$
drop out of such an analysis. In this Section we
consider the BAO distance $d_\textrm{BAO} = r_S$ as
a calibrated standard ruler and use independently
measured $h$ and $\Omega_\textrm{b} h^2$, while
keeping $N_\textrm{eff} = 3.36$ fixed, to further constrain
the cosmological parameters.

The sound horizon is calculated from first principles \cite{Eisenstein}
as follows:
\begin{equation}
r'_S = \int_0^{t_\textrm{\tiny{dec}}} {\frac{c_s dt}{a}} = 
\int_0^{a_\textrm{\tiny{dec}}} \frac{c_s da}{H_0 a^2 E(a)},
\end{equation}
where the speed of sound is
\begin{equation}
c_s = \frac{c}{\sqrt{3 (1 + 3 \rho_{b0} a / (4 \rho_{\gamma 0})}}.
\end{equation}
We can write the result for our purposes as 
\begin{equation}
r_S = 0.03389 \times A \times \left( \frac{0.30}{O_\textrm{m}} \right)^{0.255}
\end{equation} 
where
\begin{equation}
A = \left( \frac{h}{0.72} \right)^{0.489} 
  \left( \frac{0.023}{\Omega_\textrm{b} h^2} \right)^{0.098} 
  \left( \frac{3.36}{N_\textrm{eff}} \right)^{0.245}
\label{A}
\end{equation}
(we have neglected the dependence of $z_\textrm{dec} = 1090.2 \pm 0.7$ on
the cosmological parameters). We take $\Omega_\textrm{b} h^2 = 0.023 \pm 0.002$ from
Big-Bang nucleosynthesis \cite{PDG}. With the latest direct measurement $h = 0.720 \pm 0.030$
by the HUBBLE satellite \cite{hubble} we obtain $A = 1.000 \pm 0.022$.
The alternative value $h = 0.673 \pm 0.012$ is obtained from PLANK $+$ WP $+$ highL
\cite{PDG} assuming $\Omega_k = 0$ and constant $\Omega_\textrm{DE}(a)$. For this $h$
we obtain $A = 0.968 \pm 0.012$. 
The cosmological parameters that minimize the $\chi^2$ with
19 terms (17 BAO measurements from Table \ref{cs} plus $\theta_\textrm{MC}$ 
plus $A$) are presented in Table \ref{calibrated_BAO}.

From the fits to $\theta_\textrm{MC}$ plus $A$ plus each of the sets of 18, 12 or 17
BAO measurements we obtain, 
for free $\Omega_\textrm{DE}(a)$ as in Scenario 4,
$\Omega_k = 0.027 \pm 0.018$, $0.031 \pm 0.019$, 
and $0.023 \pm 0.019$ for $A = 1.000 \pm 0.022$, and  
$0.001 \pm 0.010$, $0.006 \pm 0.012$, and
$-0.004 \pm 0.010$ for $A = 0.968 \pm 0.012$.
Note that the external constraint from $A$ reduces the uncertainty on
$\Omega_k$.

\begin{table*}
\caption{\label{calibrated_BAO}
Cosmological parameters obtained from the 17 BAO measurements in Table \ref{cs}
plus $\theta_\textrm{MC}$ plus $A$ in several scenarios.
Corrections for peculiar motions are given by
Eq. (\ref{correction}).
Scenario 1 has $\Omega_\textrm{DE}(a)$ constant.
Scenario 4 has $\Omega_\textrm{DE}(a) = \Omega_\textrm{DE} \left[1 + w_1 (1 - a)\right]$.
}
\begin{ruledtabular}
\begin{tabular}{c|cccccc}
   & Scenario 1 & Scenario 1 & Scenario 4 & Scenario 4 & Scenario 4 & Scenario 4 \\
$A$ & $1.000 \pm 0.022$ & $0.968 \pm 0.012$ & $1.000 \pm 0.022$ & $1.000 \pm 0.022$ 
  & $0.968 \pm 0.012$ & $0.968 \pm 0.012$ \\
\hline
$\Omega_k$ & $0$ fixed  & $0$ fixed & $0$ fixed & $0.023 \pm 0.019$ & $0$ fixed 
  & $-0.004 \pm 0.010$ \\
$\Omega_\textrm{DE} + 2 \Omega_k$ & $0.727 \pm 0.004$ & $0.726 \pm 0.004$ & $0.721 \pm 0.006$
  & $0.716 \pm 0.007$ & $0.717 \pm 0.006$ & $0.719 \pm 0.007$ \\
$w_1$ & n.a. & n.a. & $0.313 \pm 0.213$ & $-0.049 \pm 0.377$ & $0.457 \pm 0.201$ & $0.488 \pm 0.212$ \\
$100 d_\textrm{BAO}$ & $3.44 \pm 0.02$ & $3.43 \pm 0.02$ & $3.38 \pm 0.04$
  & $3.39 \pm 0.04$ & $3.35 \pm 0.04$ & $3.36 \pm 0.04$ \\
$\chi^2/$d.f. & $19.7/17$ & $22.6/17$ & $17.4/16$ & $15.7/15$ & $16.9/16$ & $16.7/15$ \\
\end{tabular}
\end{ruledtabular}
\end{table*}

\section{Conclusions}

The results of these studies are:

(i) We define and measure BAO observables
$\hat{d}_\alpha(z, z_c)$, $\hat{d}_z(z, z_c)$, and $\hat{d}_/(z, z_c)$
that do not depend on any cosmological parameter.
From each of these observables we obtain
the BAO correlation distance $d_\textrm{BAO}$ in units of $c/H_0$
with its respective dependence on the cosmological parameters.
It is difficult to distinguish the BAO signal from the
background fluctuations due to the clustering of galaxies.
To gain confidence in the results we repeat the measurements 
many times with different galaxy selections to obtain 
different background fluctuations.
The measured BAO observables in 
Tables III and IV of Part I \cite{bh1} and Table \ref{cs}
are the main result of these studies. These measurements in 
combination with independent observations constrain cosmological
parameters.

(ii) From the BAO measurements alone we obtain the
constraint $\Omega_\textrm{DE} + 0.6 \Omega_k = 0.647 \pm 0.031$ for
constant $\Omega_\textrm{DE}(a)$, or
$0.627 \pm 0.075$ when $\Omega_\textrm{DE}(a)$ is allowed to
depend on $a$ as in Scenario 4. See Table \ref{BAO_fit} for fits
in several scenarios.

(iii) From the BAO measurements plus $\theta_{MC}$ from the CMB
we obtain the constraints
$\Omega_\textrm{DE} + 2 \Omega_\textrm{k} = 0.717 \pm 0.007$ 
and $\Omega_k = 0.060 \pm 0.052$
when $\Omega_\textrm{DE}(a)$ is allowed to vary as in Scenario 4.
See Tables \ref{BAO_thetaMC_fit_18} and \ref{BAO_thetaMC_fit_17}
for fits in several scenarios. The cosmological parameters
$h$, $\Omega_\textrm{b} h^2$ and $N_\textrm{eff}$ drop out
of this analysis.

(iv) From the BAO measurements
plus $\theta_{MC}$ from the CMB
plus external measurements of $A$ defined in Eq. (\ref{A})
we obtain
$\Omega_k = 0.023 \pm 0.019$ 
for $A = 1.000 \pm 0.022$, and
$-0.004 \pm 0.010$ for $A = 0.968 \pm 0.012$,
when $\Omega_\textrm{DE}(a)$ is allowed to vary as in Scenario 4.
For more details see Table \ref{calibrated_BAO}.
Note that the external constraint from $A$ reduces the uncertainty 
of $\Omega_k$.

(v) The data is consistent with the constraint $\Omega_k = 0$
or the constraint $\Omega_\textrm{DE}(a)$ constant, but there
is some tension when both constraints are required.
These tensions are presented in Table \ref{tension}
and in Fig. \ref{O_DE_z_all}. The measured excess of 
$\Omega_\textrm{DE}(a)$
for $0.6 < a < 0.67$ is not understood.

(vi) We note that the final consensus results of DR12 data \cite{consensus}
also show this tension, see last column in Table \ref{dr12}
which is in agreement with Fig. \ref{O_DE_z_all}.
Finally we note that the BAO measurements in this study are in
agreement with \cite{consensus}: compare Table \ref{dr12} with
Tables III or IV of Part I \cite{bh1} or Table \ref{cs}.
The two studies are complementary.

\section{Acknowledgment}
Funding for SDSS-III has been provided by the 
Alfred P. Sloan Foundation, the Participating Institutions, 
the National Science Foundation, and the 
U.S. Department of Energy Office of Science. 
The SDSS-III web site is http://www.sdss3.org/.

SDSS-III is managed by the Astrophysical Research Consortium 
for the Participating Institutions of the SDSS-III Collaboration 
including the University of Arizona, the Brazilian Participation Group, 
Brookhaven National Laboratory, Carnegie Mellon University, 
University of Florida, the French Participation Group, 
the German Participation Group, Harvard University, 
the Instituto de Astrofisica de Canarias, 
the Michigan State/Notre Dame/JINA Participation Group, 
Johns Hopkins University, Lawrence Berkeley National Laboratory, 
Max Planck Institute for Astrophysics, Max Planck Institute for Extraterrestrial Physics, 
New Mexico State University, New York University, Ohio State University, 
Pennsylvania State University, University of Portsmouth, Princeton University, 
the Spanish Participation Group, University of Tokyo, 
University of Utah, Vanderbilt University, University of Virginia, 
University of Washington, and Yale University. 

The author acknowledges the use of computing resources of
Universidad de los Andes, Bogot\'{a}, Colombia.

\end{document}